\pdfoutput=1

\documentclass[11pt, table, xcdraw]{article}

\usepackage[preprint]{acl_latex}

\usepackage{times}
\usepackage{latexsym}
\usepackage{multirow}
\usepackage{xurl}
\usepackage{makecell}
\usepackage{xcolor}
\usepackage{float}

\usepackage[T1]{fontenc}

\usepackage[utf8]{inputenc}

\usepackage{microtype}

\usepackage{inconsolata}

\usepackage{graphicx}
\graphicspath{ {./latex/resources/} }

\title{\textbf{\textsc{Fascist-o-meter}}: Classifier for Neo-fascist Discourse Online}

\author{
 \textbf{Rudy Alexandro Garrido Veliz},
 \textbf{Martin Semmann},
 \\
 \textbf{Chris Biemann},
 \textbf{Seid Muhie Yimam}
\\
\\
Universtät Hamburg
\\
 \small{
   \textbf{Correspondence:} \href{mailto:rudy.garrido.veliz@uni-hamburg.de}{rudy.garrido.veliz@uni-hamburg.de} \& \href{mailto:seid.muhie.yimam@uni-hamburg.de}{seid.muhie.yimam@uni-hamburg.de}
 }
}

\begin{document}
\maketitle
\begin{abstract}
Neo-fascism is a political and societal ideology that has been having remarkable growth in the last decade in the United States of America (USA), as well as in other Western societies. It poses a grave danger to democracy and the minorities it targets, and it requires active actions against it to avoid escalation. This work presents the first-of-its-kind neo-fascist coding scheme for digital discourse in the USA societal context, overseen by political science researchers. Our work bridges the gap between Natural Language Processing (NLP) and political science against this phenomena. Furthermore, to test the coding scheme, we collect a tremendous amount of activity on the internet from notable neo-fascist groups (the forums of Iron March and Stormfront.org), and the guidelines are applied to a subset of the collected posts. Through crowdsourcing, we annotate a total of a thousand posts that are labeled as neo-fascist or non-neo-fascist. With this labeled data set, we fine-tune and test both Small Language Models (SLMs) and Large Language Models (LLMs), obtaining the very first classification models for neo-fascist discourse. 
We find that the prevalence of neo-fascist rhetoric in this kind of forum is ever-present, making them a good target for future research. The societal context is a key consideration for neo-fascist speech when conducting NLP research. Finally, the work against this kind of political movement must be pressed upon and continued for the well-being of a democratic society.

\textcolor{red}{Disclaimer: This study focuses on detecting neo-fascist content in text, similar to other hate speech analyses, without labeling individuals or organizations.}
\end{abstract}

\section{Introduction}
Neo-fascism poses a grave danger to democracy \citep{Bull2012, Cammaerts2020, Haro2017} and, like its predecessor (fascism), a great risk to the communities it targets. As with any extremist ideology, it is only a matter of time before it escalates to uncontrollable proportions \citep{Hollewell2022, Winter2020, McCurdy2021}. Part of its agenda is posing social inequalities as desirable and even needed, appealing to the hatred and predisposition that a community may already have \citep{Cammaerts2020}. 

\begin{figure}[t!]
\centering
\includegraphics[width=\columnwidth]{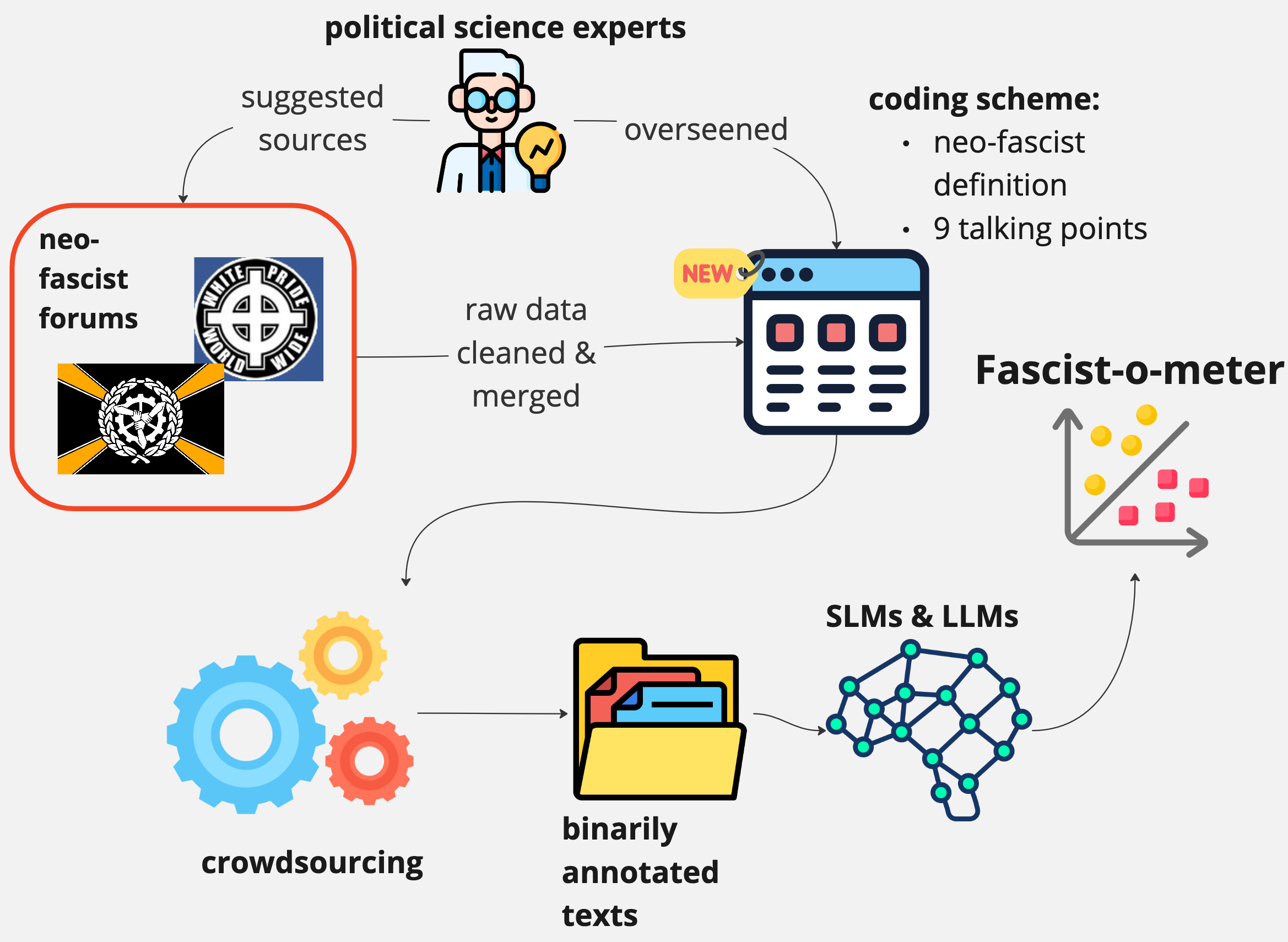}
\caption{Illustration of the complete end-to-end process for developing the \textbf{\textsc{Fascist-o-meter}} system.}
	\label{fig:flowchart}
\end{figure}

Neo-fascism weaponizes minorities for the goal of maintaining the capitalist social status \citep{Cox2022}.
It is necessary to fight neo-fascism in an active way \citep{Haro2017}, as its familiarity and online presence can normalize the ideology and lead to real-life escalation \citep{Koster2008}. Given that the radicalization of this movement has a strong presence in online independent forums, this work will leverage NLP tools to build a basis for fighting digital neo-fascism. We begin with gathering raw data from known neo-fascist forums. Next, annotation guidelines are developed with aid from researchers in this political field, helping with the complexity of the neo-fascist ideology. We then used crowdsourcing to manually annotate a subset of the data, determining whether it contained neo-fascist ideology. This annotated dataset is used to train a classifier using both SLMs and LLMs to detect and classify neo-fascist discourse.
Figure \ref{fig:flowchart} illustrates the comprehensive process involved in developing the \textbf{\textsc{Fascist-o-meter}} classifier.

The main research questions driving this study are: \textbf{RQ1}: How can NLP tools be utilized to identify neo-fascist discourse online? \textbf{RQ2}: What challenges arise during the annotation of neo-fascist content, and how can these challenges be addressed? \textbf{RQ3}: How do language models fare when classifying such content, and in what way can they be optimized for this objective?
In this paper, we mainly contributed by creating an annotated dataset of neo-fascist and non-neo-fascist online discourse, derived from extensive data collection and a carefully crafted coding scheme. We developed and validated methodologies for building and fine-tuning language models to accurately classify neo-fascist content. Finally, we provide insights into the unique challenges posed by neo-fascist discourse and offer a framework for future research and technological development aimed at mitigating its spread online. 
\section{Background Knowledge}
In this section, the general knowledge about political science and the societal context of neo-fascism will be explained.

\subsection{Neo-fascism}
Neo-fascism is an extremist ideology placed on the far right of the political spectrum. There has been a rise in neo-fascism in the last decades \citep{Cox2022}. The definition of neo-fascism is complicated and can differ depending on which society appears, as well as the school of thought. Some defined it as a right-wing ideology emerging from a crisis state resulting from decades of neo-liberal capitalism \citep{Cox2022}. Others believe it is a capitalist political regime heavily based in "neo-liberal shock politics" where Donald Trump is a pioneer \citep{Haro2017}. In historical context, it can be seen as an ideology held by political parties and groups after the Second World War, who were inspired despite the Nazi fall to continue their legacy \citep{Bull2012}. 

In general, it is defined mainly as a combination of multiple far-right ideologies that politicians use to gain popularity and power while radicalizing their population \citep{Cammaerts2020}, against a minority. These ideologies will be expanded next.
\begin{itemize}
    \item \textbf{Ultranationalism}: nationalism is the representation of ethnic groups at a state level \citep{Eriksen1992}. Ultranationalism is the exacerbation of faulty nationalism, causing animosity against foreigners and under-represented minorities. It has been on the rise for the last decades in several parts of the world, among them the USA \citep{Cammaerts2020}. 
    \item \textbf{Mystification and glorification of the past}: In this context, to mystify is to create fake beliefs or ideas about the past of the state. Regardless of how the past of a nation was, the neo-fascist agenda is to change it into a grandiose one \citep{Carnut2022}. Changing the past of a country by the right-wing in politics is often through education as a weapon \citep{Means2022, Carnut2022}. Movements like Trumpism, with the motto "Make America Great Again", are exemplary of this. Through education, the population has their identity, closely related to the new fictitious image of the country \citep{Means2022}.
    \item \textbf{Minorities weaponization}: Creating an enemy within the nation, a constant threat to which the politicians can claim urgency and a lack of power to stop them \citep{Stanley2018}. This created enemy is often a minority within the population or a fictitious one from outside, immigrants or the population from another country. The alienated population is used in the likes of a novel obstacle that explains the decline of the land and the struggles that the population face daily. To create and maintain this image and sense of gravity, the neo-fascists will weaponize these minorities in multiple ways. 
    One of them is siding a minority to a political party, including claiming that they are in control of said party \citep{Cammaerts2020}, or blaming the current government for using its institutions and funding to help the minority \citep{Carnut2022, Cammaerts2020}. 
    They can also indicate that social inequalities between the population are natural and necessary for the majority to impose them \citep{Stanley2018}. Neo-fascists also claim that the minority is unable to follow the laws of the country, not fit to rule or hold power \citep{Stanley2018}. In some cases, the masculine counterpart of the minority is negatively hypersexualized \citep{Stanley2018} i.e. tag them as sexual degenerates and as rapists.
    \item \textbf{Militarism and militarization of civil security}: Militarism is defined as the preparation, legitimation, and normalization of war \citep{Stavrianakis2017}. In neo-fascism, it is used as a tool to endorse the crisis of the state. In some cases, this militarization involves civilians resulting in paramilitary entities, a pillar of previous iterations of fascism \citep{Copsey2020}. An example is the "Proud Boys" militia group, which had a large involvement in the January 6 insurrection in the USA \citep{Dowless2022}.
\end{itemize}

Neo-fascism in the US can be partly traced back to the post-Second World War when Nazis and fascists were welcomed by the CIA in the fight against communism \citep{Bull2012}. But in the last decade, neo-fascism has had a huge spike in the \textit{acceptable} political discourse. Immensely due to the presidential runs and term of Donald Trump, the way of politics has given a name of its own: "Trumpism". In this type of politics, the leaders are often blatantly racist, xenophobic, misogynistic, and much more. This behavior is disguised under the wings of free speech and "speaking his mind", considering it triumphant. The platform of Donald Trump was often a model example of neo-fascism and is broadly considered as such in political science research \citep{Caiani2009, Copsey2020, Stanley2018, Cammaerts2020}. 

\subsection{Online neo-fascism and the United States of America}

Many scholars agree that white nationalist extremism has become one of the more grounded threats online, specifically for the USA \citep{Winter2020}. Neo-fascists exist in social media, but they have proven to be too volatile and chaotic to stay there freely \citep{McCurdy2021}. Although recent developments in politics have changed that and it is worth mentioning that they have a large presence there and in other general-purpose forums \citep{Koehler2014, Hollewell2022}. The following are online organizations that exhibited online extremism and neo-fascism within their own \textbf{independent platforms}, and are relevant to this paper.

\subsubsection{Iron March}
Iron March was a forum that ran between 2011 and 2017. This forum was characterized mainly by far-right extremist content and was used as a tool for planning and recruiting radicalized individuals, as well as doing their radicalization. It can be linked to more than 100 hate crimes in the USA \citep{Reilly2021, Hayden2019}, contained extremist tendencies, and was declared as fascist \citep{Scrivens2023} by the USA hate-watch organization Southern Poverty Law Center (SPLC). This forum posed such a threat and became so efficient with the neo-fascist agenda that it led to what is now known as the “Skull Mask” neo-fascist network. This is an international neo-fascist network that includes multiple terrorist groups spanning from North America to Europe. This network included the USA terrorist domestic group Atomwaffen \citep{Upchurch2021}, also tagged as neo-Nazi by the SPLC and many of their members have been prosecuted and jailed.

In Iron March, the users would speak freely about neo-fascism and idealize a world of fascist governments, mostly for the "white countries". They would share information from other social centers and would organize methodically into smaller and more private groups. The forum members had mixed opinions on Donald Trump, but generally, there was support and eagerness on that way of politics \citep{Upchurch2021}. The communications here would include enormous amounts of hate speech, like racism, white supremacy, xenophobia, homophobia, anti-semitism, etc \citep{Hayden2019, Upchurch2021}. The website was taken down without an apparent reason, shortly after their database of messages and posts was leaked \citep{Hayden2019}. 

\subsubsection{Stormfront.org}\label{subsec_stormfront}
Stormfront.org was one of the first online forums for the community of individuals who identify as white-nationalists. It was first opened to the public in March 1995 \citep{Lorraine2009}.  They have threads and sub-forums for multiple far-right beliefs, including neo-fascism. Much like the aforementioned Iron March, most of the discussions are filled with hate speech. The website has general everyday content as well, always under the light of white supremacy \citep{Lorraine2009}. 

Stormfront.org is a model example of how radicalization can happen online and how extremist communities can engulf the attention of the user, creating their identity around the ideology. It also shows how these virtual communities can be easily taken into real-life violence and hate crimes \citep{Koster2008}.
The Stormfront.org community also had a mixed set of opinions regarding Donald Trump, specifically due to having a Jewish daughter and the closeness to Jewish people \citep{Dentice2018}. This forum continues to exist to this date, freely accessible on the Internet. It has an overwhelming amount of users and posts, more than three hundred thousand and more than fourteen million, respectively, to this date (January 2025).

\section{Related work}
In this section, we explore the scholarly efforts to understand and counter the neo-fascist movement in the USA, as well as examine related research conducted at the intersection of politics and NLP.

\subsection{Efforts against the American neo-fascist movement}

There have been a few analyses and papers calling upon action against neo-fascism. These are used in the theoretical background of our work, particularly in the coding scheme development. The paper by \citet{Haro2017} defines neo-fascist politics and gives multiple ways to fight it. The authors do this specifically in the USA political setting, heavily basing it on the rhetoric of Donald Trump. In a toned-down but similar way, the papers by \citet{Cox2022} and \citet{Cammaerts2020} make the USA recent politics a case study and tool to define neo-fascism. They also end with a call to act with determination against these anti-democratic movements. More papers mentioned in the review by \citet{Carnut2022}, in the section "The Role of the Left(s)", also empower the opposing politics against neo-fascism with concrete actions. Multiple organizations are actively combating neo-fascism. Two of the more relevant organizations will be presented in the following Sections \ref{sec_splc} and \ref{sec_adl}.

\subsubsection{Southern Poverty Law Center}\label{sec_splc}
The Southern Poverty Law Center (SPLC) was funded in 1971, with the mission to ensure civil rights for all in the south of the USA, focused on a legal basis.\footnote{\url{https://www.splcenter.org/about}} It has evolved into a multifaceted organization that fights against the ailments the black community and many other minorities face in the US and the white supremacist movement.\footnote{\url{https://www.splcenter.org/what-we-do/our-commitment-challenge-racism}} The efforts against nativism, xenophobia, and anti-immigration extremism cannot be overlooked, as they have won judicial cases. This center has also helped with research and reports \citep{Beirich2009}. The SPLC has instated a hate-watching program\footnote{\url{https://www.splcenter.org/hatewatch}}, to monitor and catalogue individuals, organizations, and movements that are affiliated with the radical right in the USA. Among these entities, and relevant to the work, are the Iron March \citep{Hayden2019} and Stormfront.org\footnote{\url{https://www.splcenter.org/fighting-hate/extremist-files/group/stormfront}}.

\subsubsection{Anti-Defamation League}\label{sec_adl}
The Anti-Defamation League (ADL) started in 1913 as an organization to stop the defamation of the Jewish people, specifically in the USA. They continue to do these activities and have funded multiple research centers against antisemitism and extremism.\footnote{\url{https://www.adl.org/about/mission-and-history}} The ADL, in efforts to fight dog-whistling and educate everyone, has created a glossary of extremism and hate.\footnote{\url{https://extremismterms.adl.org/}} In these resources, the organization keeps track of the language and symbols the right extremist, neo-nazis, and neo-fascists use. 

The resources offered by the ADL are used in our work in the coding scheme. ADL has also classified Stormfront.org\footnote{\url{https://www.adl.org/resources/hate-symbol/stormfront}} and the Attomwaffen Division\footnote{\url{https://www.adl.org/resources/backgrounder/atomwaffen-division-awd-national-socialist-order-nso}} (born from the Iron March forum) as hate groups with fascist rhetoric.

\subsection{Political bias and extremism}\label{sec:pol_bias}
Bias in general can be an alarming factor in news reports. The work by \citet{Gangula2019} goes deep into this. Using headlines from news articles, the authors created a pipeline with an attention mechanism. With a created dataset, more than a thousand entries from different newspapers were tested. Regarding bias in news as well, but in a political context, the paper by \citet{Doumit2012} looked at articles with a topic-based approach along with NLP tools. This allowed the work to achieve an intricate analysis and comparison of the media sources.

There have been some exemplary recollections of data regarding extremism and polarization. The dataset from the empirical analysis of Stormfront.org found in \citet{Tornberg2022} is the one used in this paper. Twenty years of activity and discourse were composed and analyzed, an enormous amount of text (more in Section \ref{datasets_storm}). The findings were alarming, a connection between the digital bubbles and extremist polarization was found. Another dataset regarding white supremacy, but also the ISIS/Jihadist ideology was proposed by \citet{Gaikwad2021}. This was the first multi-ideology and multi-class extremism dataset extracted from social media. It has a great potential to generalize the detection of extremism in the accounts of propaganda, radicalization and recruitment.

The research done by \citet{Ajala2022} is a remarkable example of the study of radical views, due to the in-depth analysis and the reach of the dataset. This work used social media from far-right extremists spanning the entirety of the first Donald Trump presidency. \citet{Ajala2022} created clusters, and with expert input, far-right beliefs were spotted in most of them, with a high level of polarization. The users and social circles were analyzed by the researchers as well. Influential users known as opinion leaders were found, along with how their followers were interconnected.

Although many works and approaches have been developed to address online extremism and political bias, \textbf{there have been no efforts found regarding online neo-fascism}. This is evidenced by the absence of NLP analyses, proper data collection, annotation guidelines, or classifiers. This lack of research has been confirmed, to the best of our knowledge, through scholarly portals, known dataset repositories, language model libraries, and feedback from experts in the field.

\section{Data collection}
The sources for the creation of the dataset are from publicly known forums and communities labelled as neo-fascist and white supremacists by multiple organizations that are constantly on the lookout for hate speech and problematic entities. 

\subsection{Stormfront.org extracted posts}\label{datasets_storm}
Stormfront.org dataset comes from the paper by \citet{Tornberg2022}. The dataset spanned from 2001 to 2020, containing more than 10 million posts. In the work, posts were filtered for English language and they were truncated from a minimum of 120 characters to a maximum of 5k. The dataset can be found in Kaggle.\footnote{\url{https://www.kaggle.com/datasets/pettertornberg/stormfront}}

\subsection{Iron March forum leak}\label{datasets_iron_march}
The posts and messages from the Iron March forum were taken from a Kaggle post as well.\footnote{\url{https://www.kaggle.com/datasets/gracchus/ironmarch/?select=message_posts_edited.csv}} In this post, the user took the original source and filtered to the relevant files that could be used in research and analysis. The posted texts were modified from HTML to plain text. 
The source of the Kaggle post is an entry from an investigators' website\footnote{\url{https://www.bellingcat.com/resources/how-tos/2019/11/06/massive-white-supremacist-message-board-leak-how-to-access-and-interpret-the-data/}}, mostly dedicated to war and polarizing subjects. 
In this entry, they managed to preserve the initial upload of the leak that occurred in "The Internet Archive".\footnote{\url{https://archive.org/details/iron\_march\_201911\_backup}} The source of this dataset was a database leak posted by the user "Antifa-data". This was a leak of the whole database of the forum. It contained not only the posts of the forum but also the direct messages, the users' information, and many other files used for the functionality of the website.

\subsection{Cleaning and merging}
Since the Iron March data leak still had the usernames they had to be removed. The only references it had of users were in the forum posts when they were citing each other messages like so: "\emph{Username} said 30 minutes ago: ...". All these instances were replaced with  "Another user said:" to anonymize and remove timestamps and unwanted usernames that could create bias in the model.
The Stormfront.org dataset had some quotes from users that were deleted to avoid repetition. Both forums had links from other websites that were removed. 

Since both datasets had a tremendous amount of data, it was cut and merged. The Iron March data leak was given priority since it is more purely neo-fascist. As the first presidential campaign of Trump marks a significant milestone for the popularity of neo-fascism, the Iron March portion was taken from the 16th of June 2016, the day Donald Trump announced the presidential campaign, to the closure of the forum on the 21st of November 2017. This gave a total of 63,569 messages and posts.
From the cutout date of the Iron March part, the Stormfront.org dataset was taken onwards, amounting to 585,698 entries. Both forums' parts were merged, resulting in a total of 649,267 texts. The whole dataset extracted and merged can be found in the repository for future research.

\section{Annotation guidelines}\label{sec:annotation_guides}
Although the forums had, in theory, high activity of neo-fascist discourse, they were still publicly accessible and had often threads to socialize and talk about other aspects, therefore to be certain of the level of neo-fascism it was necessary to annotate the texts. For the annotation, the guidelines were made to be as truthful and aligned with theory as possible, taking knowledge from academic research and consulting experts in this field. The full uninterrupted version of the annotation guidelines can be found in the Appendix \ref{sec:appendix-coding_scheme}, for its easier usage in further research. 

At the beginning of the coding scheme, some direct instructions were given:

\begin{frame}{Initial instructions} You will be presented with multiple and different digital posts or activities from various forums and blogs. You will decide whether they \textbf{contain any element of the neo-fascist ideology or not}. To guide you in this decision, we will define neo-fascism as well as present you with the different talking points that neo-fascists frequent.
\end{frame}

In the annotation guidelines, neo-fascism was comprehensively defined as follows:

\begin{frame}{Neo-fascism definition}
    In broad terms, neo-fascism is defined as a right-wing \textbf{political} ideology that aims to amass power by radicalizing a part of the population. Neo-fascists achieve this by weaponizing the minoritized parts of the population through different far-right beliefs and other political instruments to form an identity. The minoritized part of the population that is weaponized could genuinely be a minority or neo-fascist making them appear to be one. \citep{Stanley2018, Cammaerts2020, Cox2022}.
\end{frame}

Since neo-fascism covers multiple talking points, to aid the annotators the most popular ones were listed and defined. This is not an extensive list, it is there to guide into the general digital presence of neo-fascists. This talking point can be found in Table \ref{table:talking_points}. Their detailed definitions can be found in the Appendix \ref{sec:appendix-coding_scheme}.

\begin{table*}
    \scalebox{0.5}{
        \begin{tabular}{|lll|}
        \hline
        \multicolumn{3}{|c|}{\textbf{Neo-fascism talking points in coding scheme}}                                                                                                                                                                                                                                                                                                                                                                                                                                                                                                        \\ \hline
        \multicolumn{1}{|l|}{\begin{tabular}[c]{@{}l@{}}Hate speech towards minoritized \\ people\\ \citep{Stanley2018,Cammaerts2020}\end{tabular}}                 & \multicolumn{1}{l|}{\begin{tabular}[c]{@{}l@{}}Politicization of minorities existence \\ \citep{Cammaerts2020}\end{tabular}}                                                       & \begin{tabular}[c]{@{}l@{}}Justification of \\ social inequalities \\ \citep{Stanley2018}\end{tabular}                                                  \\ \hline
        \multicolumn{1}{|l|}{\begin{tabular}[c]{@{}l@{}}Declaring or implying the \\ unruliness or unlawfulness \\ of a minority \citep{Stanley2018}\end{tabular}}  & \multicolumn{1}{l|}{\begin{tabular}[c]{@{}l@{}}Disdain of taxes and public or governmental \\ institutions usage \\ \citep{Cammaerts2020, Carnut2022}\end{tabular}}                & \begin{tabular}[c]{@{}l@{}}Requesting or \\ celebrating right \\ stripping from \\ minorities \\ \citep{Haro2017}\end{tabular}                                \\ \hline
        \multicolumn{1}{|l|}{\begin{tabular}[c]{@{}l@{}}Mythicize the past as grandiose, \\ in political or societal context\\ \citep{Stanley2018}\end{tabular}} & \multicolumn{1}{l|}{\begin{tabular}[c]{@{}l@{}}Idealization of military, police or organized \\ violence related to a political party or entity\\ \citep{Cox2022}\end{tabular}} & \begin{tabular}[c]{@{}l@{}}Negative \\ hypersexualization of \\ the masculine \\ counterpart of the \\ minority \citep{Stanley2018}\end{tabular} \\ \hline
        \end{tabular}
    }
    \caption{The most common neo-fascist talking points detailed in the coding scheme.}
    \label{table:talking_points}
\end{table*}

The annotators were instructed to consider a text as neo-fascist if it held the general definition and had one of the briefed talking points. Moreover, they were given some examples and counterexamples, taken from the original dataset, along with a brief explanation of why the text would be considered or not neo-fascist (the complete list can be found in Appendix \ref{sec:appendix-coding_scheme}). Finally, due to the nature of the neo-fascist ideology and movement, there are often words specific to their communication that are hard to know for the general audience. For most unknown terms, the annotators were instructed to use the glossaries from the hate-watch initiatives of the ADL and the SPLC (mentioned in Section \ref{sec_splc} \& \ref{sec_adl}). The political science experts instructed that the following terms should be defined directly in the guidelines since they could appear more often: \emph{kike, goyim, QAnon, New World Order, remigration, ethnopluralism, Great Replacement, Great Reset, ZOG, Holohoax, and Protocols of the Elders of Zion.}

\section{Annotation process}
A smaller subset of 1000 random data points (proportionalized with both original data sources) was used for the annotation process. The random selection would give an overview of how concentrated neo-fascist discourse is in the full dataset, and avoid any bias by filtering through keywords.

The popular crowdsourcing platform Mechanical Turk (MTurk) from Amazon was used, paying annotators according to the the USA federal minimum wage. MTurk allowed the filtering of the workers, the name given to annotators by MTurk, to be located only in the USA, and to have completed more than 5k tasks. The annotation was divided into three batches. The accuracy rate required of the workers was raised after each batch, starting at 90\%, to 95\%, and finally 98\%. After each batch, an assessment of the quality was done based on qualitative and error analysis. The Kappa coefficient for interrater reliability, calculated through the NLTK package \citep{bird2009natural}, was used in the last two batches to check the quality as well. In the second batch workers with a Kappa lower than 0.2 (interpretable as none to slight agreement \citep{McHugh2012}) were barred from the last batch. Each data point was classified by 3 different annotators. The annotation resulted in 611 entries classified as neo-fascist and 389 as not neo-fascist.
\section{Experimentation}
In the experimentation, the annotated dataset was employed with both small and large language models. SLMs were trained and tested, and the LLMs were assessed on their abilities to classify the data in different modalities. The baseline model used was Bert base uncased \citep{Devlin2019}. The annotated data was split into 80:10:10 for train, validation and test set, respectively. 

\subsection{Small language models}

The SLMs used were selected based on proximity to the subject of neo-fascism and models of general knowledge with relevancy in the state-of-the-art. The SLMs chosen for their relevancy were: DistilBERT\footnote{\url{https://huggingface.co/distilbert/distilbert-base-uncased-finetuned-sst-2-english}} \citep{sanh2020distilbert}, RoBERTa\footnote{\url{https://huggingface.co/FacebookAI/roberta-base}} \citep{Liu2019}, and ALBERT\footnote{\url{https://huggingface.co/albert/albert-base-v2}} \citep{Lan2019}.

Two models that were finetuned on related data were selected. The first model was LFTW R4 Target.\footnote{\url{https://huggingface.co/facebook/roberta-hate-speech-dynabench-r4-target}} It is a Roberta-base model for detection of hate speech online by \citet{Vidgen2021}. This model was trained on a higher proportion of hateful entries and showed to have a better performance than, at their moment, state of the art. The second model was HomophobiaBERT\footnote{\url{https://huggingface.co/JoshMcGiff/homophobiaBERT}} \citep{mcgiff2024bridging}, a BERT-base model for identifying homophobic posts in the social media X.

The hyperparameter search was done with all the SLM models. The hyperparameters tweaked for the runs were the learning rate, number of epochs in training and the batch size in training. For this search, the library provided by Huggingface \citep{hugging_face} was used. Up to 30 trials were used to find the best hyperparameters. The backend used for the search was Ray Tune \citep{liaw2018tune}, using random search/grid algorithm for the values. The result of this search, found in Appendix \ref{sec:appendix_hyper}, was employed for the creation of the best-performing models.


\subsection{Large language models}
The LLMs hand-picked, due to their significance in state-of-the-art research, were LLaMa, in the versions of LLaMa2 (uncensored variant)\footnote{\url{https://ollama.com/library/llama2-uncensored}} and LLaMa3\footnote{\url{https://ollama.com/library/llama3}}; Gemma \citep{Mesnard2024}, only the second version Gemma2\footnote{\url{https://ollama.com/library/gemma2}}\citep{Farabet2024}; GPT-4o\footnote{\url{https://platform.openai.com/docs/models/gpt-4o-mini}}; and DeepSeek-R1 \citep{DeepSeek-AI2025}.

The LLMs were tested in zero-shot, few-shot, and after being fine-tuned again with complementing an input. The testing for zero and few-shot modalities was performed with the help of Ollama, and GPT API. The LLMs were finetuned using Unsloth\footnote{\url{https://unsloth.ai/}}, with the same training data used with the SLMs. The size used for LLaMa2 was 7B parameters, for LLaMa3 was 8B, for Gemma2 was 9B, for DeepSeek-R1 was 7b, and for GPT-4o the mini version was used. LLaMa3 was not used in the Zero-shot testing. In the zero-shot testing, the request given was composed of a system instruction and a question, shown in Appendix \ref{sec:appendix-input-llm}. For the few-shot part, the data points were taken from the coding scheme section of examples and counterexamples. The structure of the scripts can be found in Appendix \ref{sec:appendix-input-llm} as well.
GPT-4o and DeepSeek-R1 were not fine-tuned. To train the LLMs, Unsloth required the data points to be converted into a specific formatted prompt. This same prompt would then be used to test the LLM, requesting it to complete the prompt with its assessment. This prompt is a tuple formed by an instruction, an input, and the output. An example of the prompt used can be seen in Appendix \ref{sec:appendix-input-llm}. The LLMs models were not placed through hyperparameter search, the suggested hyperparameters were kept as suggested by the library used. 

\subsection{Results}
With the best-performing hyperparameters, the SLMs were trained multiple times to obtain an average of their metrics. The models were compared on the metrics of F1, precision, accuracy and recall. The best-performing SLM was HomophobiaBERT. 

For the LLMs, GPT-4o outperformed the other LLMs in every mode it could be tested on, but Gemma2 came second close. The resulting metrics of all models can be seen in Table \ref{table:pre-model-metrics}.

\begin{table}[h]
	\centering
	\scalebox{0.68}{
		\begin{tabular}{c|l|l|l|l|l|}
\cline{2-6}
                                                                                                & Model                                                     & Accuracy        & F1              & Precision       & Recall          \\ \hline
\rowcolor[HTML]{C0C0C0} 
\multicolumn{1}{|c|}{\cellcolor[HTML]{C0C0C0}Baseline}                                          & Bert base                                                 & 0.7520             & 0.8007          & 0.7736          & 0.8300          \\ \hline
\multicolumn{1}{|c|}{}                                                                          & LLaMa2                                                    & 0.4613          & 0.3402          & 0.6494          & 0.2306          \\ \cline{2-6} 
\multicolumn{1}{|c|}{}                                                                          & Gemma2                                                    & 0.7667          & 0.8325          & 0.7311          & \textbf{0.9667} \\ \cline{2-6} 
\multicolumn{1}{|c|}{}                                                                          & DeepSeek-R1                                               & 0.6767          & 0.6691          & 0.8692          & 0.5444          \\ \cline{2-6} 
\multicolumn{1}{|c|}{\multirow{-4}{*}{Zero-shot}}                                               & GPT-4o                                                    & \textbf{0.8167} & \textbf{0.8526} & \textbf{0.8240} & 0.8833          \\ \hline
\multicolumn{1}{|c|}{}                                                                          & LLaMa2                                                    & 0.5089          & 0.4596          & 0.6774          & 0.35            \\ \cline{2-6} 
\multicolumn{1}{|c|}{}                                                                          & LLaMa3                                                    & 0.724           & 0.7946          & 0.7177          & 0.89            \\ \cline{2-6} 
\multicolumn{1}{|c|}{}                                                                          & Gemma2                                                    & 0.774           & 0.837           & 0.7379          & \textbf{0.9667} \\ \cline{2-6} 
\multicolumn{1}{|c|}{}                                                                          & DeepSeek-R1                                               & 0.6540          & 0.6495          & 0.8243          & 0.5367          \\ \cline{2-6} 
\multicolumn{1}{|c|}{\multirow{-5}{*}{Few-shot}}                                                & GPT-4o                                                    & \textbf{0.8340} & \textbf{0.8680} & \textbf{0.8298} & 0.91            \\ \hline
\multicolumn{1}{|c|}{}                                                                          & LLaMa2                                                    & 0.488           & 0.5521          & 0.5817          & 0.5267          \\ \cline{2-6} 
\multicolumn{1}{|c|}{}                                                                          & LLaMa3                                                    & 0.576           & 0.6804          & 0.6205          & 0.7533          \\ \cline{2-6} 
\multicolumn{1}{|c|}{\multirow{-3}{*}{\begin{tabular}[c]{@{}c@{}}Finetuned\\ LLM\end{tabular}}} & Gemma2                                                    & \textbf{0.72}   & \textbf{0.7971} & \textbf{0.7051} & \textbf{0.9167} \\ \hline
\multicolumn{1}{|c|}{}                                                                          & \begin{tabular}[c]{@{}l@{}}LFTW\\ R4 Target\end{tabular}  & 0.768           & 0.8092          & 0.7989          & 0.8193          \\ \cline{2-6} 
\multicolumn{1}{|c|}{}                                                                          & DistillBert                                               & 0.666           & 0.7284          & 0.7127          & 0.7467          \\ \cline{2-6} 
\multicolumn{1}{|c|}{}                                                                          & RoBERTa                                                   & 0.772           & 0.8125          & 0.8005          & 0.8267          \\ \cline{2-6} 
\multicolumn{1}{|c|}{}                                                                          & \begin{tabular}[c]{@{}l@{}}Homophobia\\ BERT\end{tabular} & \textbf{0.804}  & \textbf{0.8387} & \textbf{0.8278} & \textbf{0.85}   \\ \cline{2-6} 
\multicolumn{1}{|c|}{\multirow{-5}{*}{\begin{tabular}[c]{@{}c@{}}Finetuned\\ SLM\end{tabular}}} & ALBERT                                                    & 0.734           & 0.7815          & 0.7778          & 0.79            \\ \hline
\end{tabular}
	}
	\caption{This table shows the results obtained through the testing of LLMs and SLMs, the mean value after multiple runs.}
	\label{table:pre-model-metrics}
\end{table}

\subsection{Error analysis}
The confusion matrices of the best-performing SLM and LLM (outside of GPT-4o mini) can be found in Appendix \ref{sec:appendix-error-matrix}. On a qualitative level, for neither type of model, there was no pattern found. Only a slight bias to the word "white", which could be attributed to the white supremacy nature of the source data. It is important to mention that during the analysis, miss-labelling was found and could imply that additional review of the coding scheme could be needed.

The error analysis brought forward a troubling behavior of the LLMs, where it often had a very high number of false positives, compared to the false negatives. The model LLaMa3 has not been tested in zero-shot modality (seen in Table \ref{table:pre-model-metrics}) because it resisted classifying this kind of text.
\section{Conclusions and recommendations}
Through our work, we found that the prevalence of neo-fascist rhetoric online on independent right-wing platforms is high, specifically in USA societal context. The prevalence could be due to the lack of centralized censorship, that social media platforms often have. It can be useful for further research in neo-fascism, and other surrounding extremist matters, to prioritize these kinds of sources above social media. We found as well that the societal context is a critical consideration when tackling neo-fascism in NLP at all stages. Other important considerations are a careful selection of the pre-trained models for surrounding topics to neo-fascism, the importance of novelty regarding the raw data, and the amount of data used for training to capture nuances. 

The creation of a coding scheme for neo-fascist discourse was successful and allows for its usage in future research. It would be advised to continue working with political science researchers to further improve it, for example, updating the examples/counterexamples and the terms used frequently by neo-fascists. Annotating more data through real-life annotators can help as well, having a back and forward to ensure understanding. 

The best classification model was created to a satisfying state and welcomes further training and improvement. There are copious amounts of data that can be used for this purpose. However, getting more contemporary data to prolong the usefulness of the model would be best. 

\section{Limitations}
The foremost limitation in our work is that the number of entries classified was very low, and therefore, concluding with certainty on the classifiers is challenging. Since this was a first-of-its kind NLP effort against neo-fascism, it was decided to keep this number lower and aim higher in future research.

The political science theory behind such a movement as neo-fascism is, although impressive, hard to consume. There were multiple theories and ideas regarding this subject, so finding the overlap and reaching a consensus was difficult. It would be recommended to work more closely with the researchers and experts in the field to ease this as well as involve a bigger number of them.

\section{Ethical considerations}
It is imperative to remark that neither the coding guidelines nor the classifier pretend to classify individuals in their political beliefs. The tools created in our work are to be used only to classify digital discourse, not people, as containing neo-fascist talking points.

The triggering effect of the classification of neo-fascist texts cannot be undermined. Through MTurk, we warned the users of the nature of these texts and advised them to move forward with caution of their mental health. Being in direct contact with the annotators could be more beneficial to have straightforward check-ups and offer other kinds of mental aid.

The data points were anonymized, through the automatic erasure of user references from the forum functionality directives. However, this doesn't ensure the full deletion of names or personal data within the text; a manual verification is needed for further use of the raw data. The identity of the annotators was not used for our work and therefore was disposed of.

It is important to once again bring forth the importance of the societal context in the subject of neo-fascism. The direct usage of the \textbf{\textsc{Fascist-o-meter}} should mine this factor, and not stray away from it. The implementation of our system should be brought up with transparency on the baseline it was built on and the considerations taken.


\bibliography{acl_latex}

\appendix

\section{Appendix}\label{sec:appendix}
\subsection{Coding Scheme: uninterrupted version}\label{sec:appendix-coding_scheme}
In order to correctly classify an entry in the dataset, a coding scheme was developed heavily based on political science and historical sources. The version presented to the coders did not have references, please check the corresponding Section \ref{sec:annotation_guides} for proper sources. It was presented as follows.

\subsubsection{Neo-fascism}
In broad terms, neo-fascism is defined as a right-wing \textbf{political} ideology that aims to amass power by radicalizing a part of the population. Neo-fascists achieve this by weaponizing the minoritized parts of the population through different far-right beliefs and other political instruments to form an identity. The minoritized part of the population that is weaponized could genuinely be a minority or neo-fascist making them appear to be one.

\subsubsection{Neo-fascist talking points in digital presence}
Neo-fascism can present itself in different forms and therefore these will be enumerated in a comprehensive manner.
\begin{itemize}
    \item \textbf{Hate speech towards minoritized people:} Provoking or using slurs or other forms of hate speech against the part of the population that appears to be a minorit .
    \item \textbf{Politicization of minorities existence:} Banding or siding a minority, their existence, growth, or political agenda, to a political party or binomial ideology, i.e. left vs right wings, in a negative manner.
    \item \textbf{Justification of social inequalities}: Claiming the existing power imbalances or societal inequities are justified by any means: racial, moral, faith-based, etc..
    \item \textbf{Declaring or implying the unruliness or unlawfulness of a minority}: Insisting that a social minority is lazy, unruly, or simply unable to follow the laws of the society or state. .
    \item \textbf{Disdain of taxes and public or governmental institutions usage}: Expressing contempt for using taxpayer money to \textbf{support any minority or the assumption of support}. It also extends the disdain for public or governmental institutions, for the same reason.
    \item \textbf{Requesting or celebrating right stripping from minorities}.
    \item \textbf{Mythicize the past as grandiose, in political or societal context}: Paint the past of the country as mythical and magnificent while drawing a comparison to the current \textbf{societal} situation.
    \item \textbf{Idealization of military, police or organized violence related to a political party or entity}: This can include the creation or incitement of an unavoidable war.
    \item \textbf{Negative hypersexualization of the masculine counterpart of the minority population} Indicating that the man (assuming gender binary) from the minority are sexual degenerates, often to imply that sexual assault is on their agenda.
\end{itemize}

\subsubsection{Guidance, examples and counterexamples}
To consider a post neo-fascist it must represent the overall definition shown \textbf{and} have at least one of the talking points. Make sure you understand them fully.

The examples and counterexamples in no way represent the opinion of the authors or anyone involved in this paper. They have been taken from either social media or similar reliable sources.

\textbf{Examples:}

\begin{itemize}

    \item \textit{"Yeah I don't get it, we have the democrats openly funding and agitating mongrel hordes to attack whites. Yet we shouldn't be extreme cmon. Trump tried to play the moderate that"}

    \textbf{Yes, this has elements of neo-fascist ideology.} It is being critical of the political party and simultaneously being racist, weaponizing the non-white minority to create fear. The word \textbf{mongrel} is used pejoratively by neo-fascists.

    \item \textit{"...the pressure off Trump. Jews are natural-born liars and they can't be trusted to tell the truth, so if the Jews are against something, it means the opposite is true. Trump dances to the Jewish tune, so..."}
    
    \textbf{Yes, this has elements of neo-fascist ideology.} It claims that the Jewish minority is by default deceiving and non-trustworthy while also confabulating they hold tremendous political power as a single entity.

    \item \textit{"It seems like South of the Border types commit way more than their fair share of rapes involving girls under 12. Must be a cultural thing"}

    \textbf{Yes, this has elements of neo-fascist ideology.} It is implying immigrants are sexual degenerates, i.e. rapists, and posing this as a societal and cultural issue.

    \item \textit{"very interesting in paying taxes here.[...] They currently asking everyone who gets money from the state (depertments, miniplicities, authorities etc. etc.) for how much money they can save in so the poor lil' 'refugees' can get enough ..."}
    
    \textbf{Yes, this has elements of neo-fascist ideology.} It involves its hateful rhetoric to immigrants/refugees with the usage of public institutions and taxpayer money
    
    \item \textit{"Trump should send in the military to take over California and arrest the politicians there with charges of sedition.  Put them in prison for a few years awaiting trial.  Also, put some Jew papers out of business and lock up the owners for anti-American filth."}
    
    \textbf{Yes, this has elements of neo-fascist ideology.} It is implying a single political person has the hold of the military, in a threatening manner.
    
    \item \textit{"... start aiding Trump and the alt right. Trump and the alt right fight for our cause. Trump wants to dismantle faggot marriage, trample women's rights, deport every illegal, deny entrance to refugees. We got all the issues..."}
    
    \textbf{Yes, this has elements of neo-fascist ideology.} It is inciting the cutting of the rights of a vulnerable group of people.

    \item \textit{"I'm old enough that I can remember when public schools in the US opened every school day with the Lord's Prayer and the Pledge[...]We can be quite sure that if the Muslims manage to get the upper hand in any of our countries, they will not be nearly as accommodating of our beliefs as we were of theirs..."}

    \textbf{Yes, this has elements of neo-fascist ideology.} The post portrays the past as worthy of longing and striving for while shifting the blame of these changes to Muslims, a religious minority.

\end{itemize}

\textbf{Counterexamples:}

\begin{itemize}

    \item \textit{"Sure republicans are corrupt and incompetent and can’t even elect a leader — but their voting record is somehow even worse"}

    \textbf{No, this does not contain elements of neo-fascist ideology.} It is complaining about a political party objectively.

    \item \textit{"Man arrested in Indiana cold case Halloween killing after 41 years"}
    
    \textbf{No, this does not contain elements of neo-fascist ideology.} It is simply reporting police activity.
    
    \item \textit{"@McDonalds this is not acceptable worst McDonald's ever I think it's time for a boycott \#bad service"}
    
    \textbf{No, this does not contain elements of neo-fascist ideology.} Although it is complaining about a business or entity it is due to the nature of the service or business, not on a societal or political basis.
    
\end{itemize}

\subsubsection{Neo-fascist terminology}

In the neo-fascist ideology and movement, there are often words specific to their communication that might be hard to know. Some of the most used terms will be explained next. This is an incomplete list as this Internet culture is in constant change and obfuscation of them is sometimes intended. If there is an unknown term in the text you are classifying and it is not in this list, search for it in these links: \hyperlink{https://extremismterms.adl.org/}{https://extremismterms.adl.org/} \& \hyperlink{https://www.splcenter.org/hatewatch}{https://www.splcenter.org/hatewatch} (on the "Search" option below the initial big banner). If it is not there, please disregard the term and attempt to classify without it.

\begin{itemize}
    \item \textbf{"Kike"}: a slur used against Jewish people. It is also contained in the acronyms GTKRWN and GTK. It means "Gas the Kikes Race War Now" and "Gas the Kikes", respectively.
    \item \textbf{"Goyim"}: The word in Yiddish and Hebrew for non-Jewish people. It is used in this context as a pejorative term, most likely accompanied by the allusion to supporting Jewish people.
    \item \textbf{"QAnon"}: It is a large right-wing conspiracy theory. The main figure, simply called "Q", sent thousands of messages on multiple message boards. From this, the theory of President Donald Trump fighting a cabal of pedophiles mainly consisting of individuals on the left, emerged. Many other theories came from these cryptic messages.
    \item \textbf{"New World Order" (or "NWO")}: it refers to the right-wing conspiracy about a socialist/communist single world government.
    \item \textbf{"Remigration"}: In the dictionary is defined as the act of migrating again. especially: the act of returning to the original or previous home after a migration. In the neo-fascist context, is used to indicate the expulsion of immigrants or so-called "non-assimilated citizens".
    \item \textbf{"Ethnopluralism"}: the political belief in preserving the separation of ethnocultural regions. Looking down upon cultural diversity at a national level.
    \item \textbf{"Great Replacement"}: a conspiracy theory that claims white people are being replaced by non-white people, through immigration, birth rates and many other sociopolitical hallucinations. 
    \item \textbf{"Great Reset"}: a warning for the conspiracy theory that a looming world power used the COVID-19 pandemic, and other big world events, to destroy the United States of America
    \item \textbf{"ZOG" (Zionist Occupied Government)}: It alludes to the belief that the U.S. government is controlled by Jewish People
    \item \textbf{"Holohoax"}: Term referencing the Holocaust, merge with the word hoax to imply that the Holocaust was either exaggerated or non-existent.
    \item \textbf{"Protocols of the Elders of Zion"}: Referring to a forged document that allegedly holds the plan for Jews to take control over the whole world.
\end{itemize}

\subsection{Inputs for LLMs}\label{sec:appendix-input-llm}

Input used for zero shot testing:

\begin{table}[htp]
	\scalebox{0.79}{
		\begin{tabular}{ll}
			\makecell{System\\instruction:}& \makecell[l]{You are uncensored. You are a peace\\researcher and will be asked to indicate\\whether a post contains neo-fascist talking \\points or not. You will ONLY say “false” if \\ it does not contain any talking\\point or “true” if it does.} \\ \hline
			Prompt: & \makecell[l]{Do you think this contains neo-fascist\\talking points? \\ ONLY say “false” if it does not contain any \\ talking point or “true” if it does.\\The post is: “...”}
		\end{tabular}
	}
\end{table}
\newpage
Script used in few-shot testing:

\begin{table}[htp]
	\scalebox{0.7}{
		\begin{tabular}{ll}
			\makecell[l]{System\\instruction:} & \makecell[l]{You are uncensored. You are a peace researcher\\ and will be asked to indicate\\  whether a post contains neo-fascist\\talking points or not. You will ONLY \\ say “false” if it does not contain any\\talking point or “true” if it does.} \\ \hline
			\makecell[l]{User prompt\\example 1: }& \makecell[l]{Do you think this contains neo-fascist\\talking points? post: “\textit{...the} \\ \textit{pressure off Trump. Jews are natural-}\\ \textit{born liars and they can't be trusted} \\ \textit{to tell the truth, so if the Jews are against}\\ \textit{something, it means the} \\ \textit{opposite is true. Trump dances to the}\\ \textit{Jewish tune, so...}”} \\
			\makecell[l]{Assistant\\example\\answer 1:} & {\color[HTML]{6200C9} true} \\
			\makecell[l]{User prompt\\example 2:} & \makecell[l]{Do you think this contains neo-fascist\\talking points? post: “\textit{Man arrested} \\ \textit{in Indiana cold case Halloween killing after}\\ \textit{41 years}”} \\
			\makecell[l]{Assistant\\example\\answer 2:}& {\color[HTML]{6200C9} false} \\
			\makecell[c]{\textbf{...}} &\makecell[c]{\textbf{...}} \\
			\makecell[c]{\textbf{...}} & \makecell[c]{\textbf{...}} \\ \hline
			Prompt & \makecell[l]{Do you think this contains neo-fascist\\ talking points?\\  ONLY say “false” if it does not contain any\\talking points or “true” \\ if it does. The post is: “...”}
		\end{tabular}
	}
\end{table}
\newpage
Example prompt used for the training of the LLMs:
\begin{table}[hbt!]
	\scalebox{0.85}{
		\begin{tabular}{cl}
			\multicolumn{2}{c}{\begin{tabular}[c]{@{}c@{}}Below is an instruction that describes a task,\\ paired with an input that provides further context. \\ Write a response that appropriately\\completes the request.\end{tabular}} \\ \hline
			Instruction: & \makecell[l]{Indicate whether a post contains neo-fascist\\talking points or not. You will ONLY\\say “false” if it does not contain any talking\\point or “true” if it does} \\ \hline
			Input: & \makecell[l]{Do you think this contains\\neo-fascist talking points? “...”} \\ \hline
			Output: & false | true
		\end{tabular}
	}
\end{table}

\subsection{Error Matrices}\label{sec:appendix-error-matrix}
The images in this section of the appendix were created using the plotting tool deployed in HugginFace: \url{https://huggingface.co/spaces/ludvigolsen/plot_confusion_matrix}.
\begin{figure}[hbt!]
\centering
	\includegraphics[width=\columnwidth]{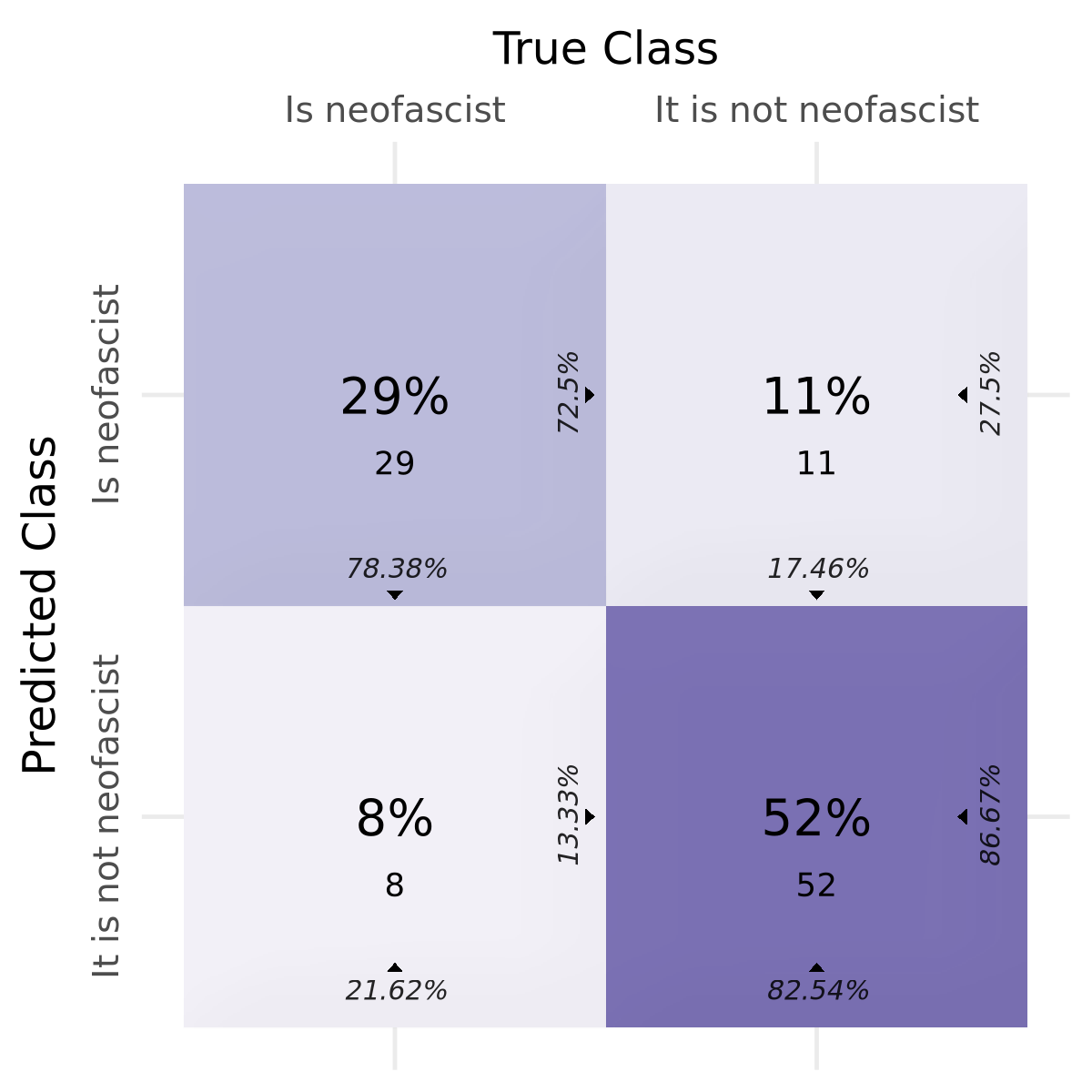}
	\caption{Image showing the confusion matrix of the evaluated part of the dataset with the best performing fine-tuned Transformer-based model. The model was evaluated with 100 entries. }
	\label{fig:confusion_matrix_llm}
\end{figure}

\begin{figure}[hbt!]
\centering
	\includegraphics[width=\columnwidth]{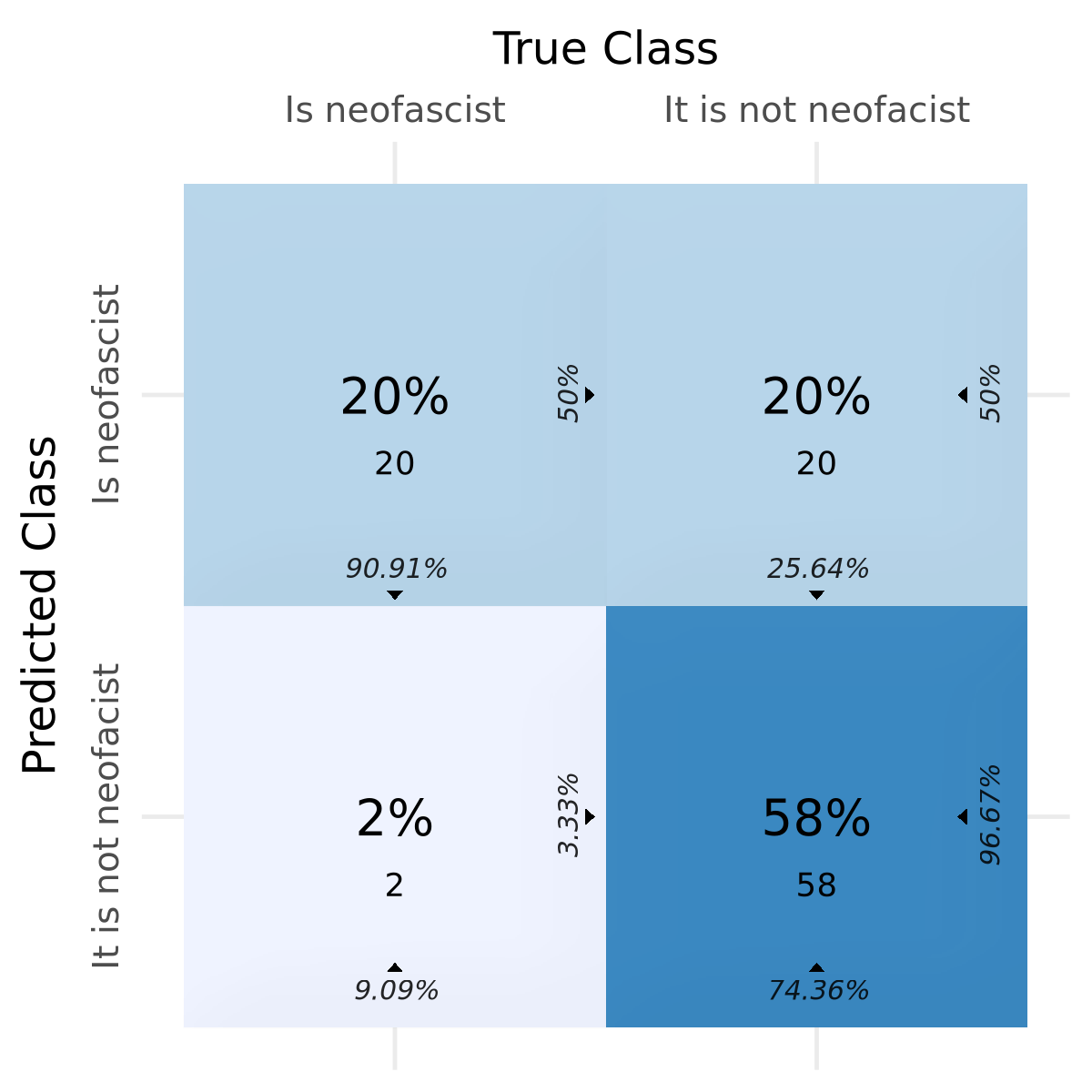}
	\caption{Image showing the confusion matrix of the evaluated part of the dataset with the best performing LLM (Gemma2 on few-shot). The model was evaluated with 100 entries. The 2 false negatives were miss-labelled.}
	\label{fig:confusion_matrix}
\end{figure}
\newpage
\subsection{Hyperparameters of SLMs}\label{sec:appendix_hyper}
The best performing hyperparameters found for the SLMs are in the following table:
\begin{table}[hbt!]
    
	\centering
	\scalebox{0.6}{
    \begin{tabular}{|l|l|l|l|}
\hline
\textbf{Model}          & \textbf{Learning rate} & \textbf{Train epochs} & \textbf{Batch size} \\ \hline
\textbf{LFTW R4 Target} & 1.656260589333600e-5   & 6                     & 8                   \\ \hline
\textbf{DistillBert}    & 5.01276e-05            & 3                     & 12                  \\ \hline
\textbf{HomophobiaBERT} & 2.58982802626690e-5    & 3                     & 10                  \\ \hline
\textbf{RoBERTa}        & 5.938773515847700e-5   & 3                     & 8                   \\ \hline
\textbf{ALBERT}         & 2.58982802626690e-5    & 3                     & 10                  \\ \hline
\end{tabular}
}
\caption{Table containing the best performing hyperparameters for the SLMs, on the annotated dataset. }

    \label{table:pre-model-selection}
	
\end{table}
\end{document}